\documentclass{ws-ijbc}
\usepackage{ws-rotating}
\usepackage{amssymb}
\usepackage{graphicx}
\begin{document}

\catchline{}{}{}{}{} % Publisher's Area please ignore

\markboth{C. G. Antonopoulos et al.}{Weak Chaos Detection in the Fermi-Pasta-Ulam-$\alpha$ System Using $q$-Gaussian Statistics}

\title{(This paper is for the Special Issue edited by Prof. Gregoire Nicolis, Prof. Marko Robnik, Dr. Vassilis Rothos and Dr. Haris Skokos)\\\vspace{1cm}WEAK CHAOS DETECTION IN THE FERMI-PASTA-ULAM-$\alpha$ SYSTEM USING $q$-GAUSSIAN STATISTICS}

\author{CHRIS G. ANTONOPOULOS}
\address{Department of Automation and\\High Performance Computing Systems - Programming \& Algorithms Lab (HPCS Lab), Technological Educational Institute of Messolonghi, Nea Ktiria, 30200\\Messolonghi, Greece\\chantonopoulos@teimes.gr}

\author{HELEN CHRISTODOULIDI}
\address{Dipartimento di Matematica Pura e Applicata, Universit\'{a} di Padova, Via Trieste, 63 - 35121\\Padova, Italy\\echrist@math.unipd.it}

\maketitle

\begin{history}
\received{(to be inserted by publisher)}
\end{history}

\begin{abstract}
We study numerically statistical distributions of sums of orbit coordinates, viewed as independent random variables in the spirit of the Central Limit Theorem, in weakly chaotic regimes associated with the excitation of the first ($k=1$) and last ($k=N$) linear normal modes of the Fermi-Pasta-Ulam-$\alpha$ system under fixed boundary conditions. We show that at low energies ($E=0.19$), when the $k=1$ linear mode is excited, chaotic diffusion occurs characterized by distributions that are well approximated for long times ($t>10^9$) by a $q$-Gaussian Quasi-Stationary State (QSS) with $q\approx1.4$. On the other hand, when the $k=N$ mode is excited at the same energy, diffusive phenomena are \textit{absent} and the motion is quasi-periodic. In fact, as the energy increases to $E=0.3$, the distributions in the former case pass through \textit{shorter} $q$-Gaussian states and tend rapidly to a Gaussian (i.e. $q\rightarrow 1$) where equipartition sets in, while in the latter we need to reach to $E=4$ to see a \textit{sudden transition} to Gaussian statistics, without any passage through an intermediate QSS. This may be explained by different energy localization properties and recurrence phenomena in the two cases, supporting the view that when the energy is placed in the first mode weak chaos and ``sticky'' dynamics lead to a more gradual process of energy sharing, while strong chaos and equipartition appear abruptly when only the last mode is initially excited.
\end{abstract}


%\begin{multicols}{2}

%%%%%%%%%%%%%%%%%%%%%%%%%%%%%%%%%%%%%%%%%%%%%%%%%%%%%%%%%%%%%%%%%%%%%%%%%%%%
%%%%%%%%%%%%%%%%%%%%%%%%%%%%%%%%%%%%%%%%%%%%%%%%%%%%%%%%%%%%%%%%%%%%%%%%%%%%
\section{Introduction}
%%%%%%%%%%%%%%%%%%%%%%%%%%%%%%%%%%%%%%%%%%%%%%%%%%%%%%%%%%%%%%%%%%%%%%%%%%%%
%%%%%%%%%%%%%%%%%%%%%%%%%%%%%%%%%%%%%%%%%%%%%%%%%%%%%%%%%%%%%%%%%%%%%%%%%%%%

Probability density functions (pdfs) of chaotic trajectories of dynamical systems have been studied for a long time by many authors, with the purpose of understanding the transition from deterministic to stochastic dynamics \cite{Anosov1967,Arnold1967,Sinai1972,Pesin1976,Pesin1977,Ruelle1979,katok1980,Ruelle1980,Ruelle1982,Eckmann1985,katok1985}. The fundamental question in this regard concerns the existence of an appropriate invariant probability density (or ergodic measure) which characterizes chaotic motion in phase space regions where solutions generically exhibit exponential divergence of nearby trajectories. If such an invariant measure can be found for almost all initial conditions (i.e. except for a set of measure zero), one has a firm basis for studying the system from a Statistical Mechanics point of view.

If this invariant measure turns out to be a continuous and sufficiently smooth function of the phase space coordinates, one can invoke the Boltzmann-Gibbs (BG) microcanonical ensemble and attempt to evaluate all relevant quantities of equilibrium Statistical Mechanics, like partition function, free energy, entropy, etc. On the other hand, if the measure is absolutely continuous (as e.g. in the case of the so-called Axiom A dynamical systems), one might still be able to use the formalism of modern ergodic theory and SRB measures (after Sinai, Ruelle and Bowen) to study the statistical properties of the model \cite{Eckmann1985}.

Thus, viewing the values of one (or a linear combination) of components of a chaotic solution at discrete times $t_n,\;n=1,\ldots,\mathcal{N}$ as realizations of $\mathcal{N}$ independent and identically distributed random variables $X_n$ and calculating the distribution of their sums, in the sense of the Central Limit Theorem (CLT) \cite{Rice1995}, one expects to find a \textit{Gaussian} pdf, whose mean and variance are those of the $X_n$'s. This is indeed what happens for many chaotic dynamical systems studied to date which are \textit{ergodic}, i.e. almost all their orbits (except for a set of measure zero) pass arbitrarily close to any point of the constant energy manifold, after sufficiently long times. In these cases, at least one Lyapunov exponent is positive, stable periodic orbits are absent and the constant energy manifold is uniformly covered by chaotic orbits, for all but a (Lebesgue) measure zero set of initial conditions.

It is also important, however, to study ``small size'' chaotic regions of Hamiltonian systems, at energies where the maximum (positive) Lyapunov exponent is ``small'' and stable periodic orbits exist with islands of invariant tori and sets of cantori, which occupy a positive measure subset of the energy manifold. In such regimes of ``weak chaos'', a great number of orbits stick for long times to the boundaries of these islands and chaotic trajectories diffuse slowly through multiple connected regions in a highly non-uniform way \cite{Aizawa1984,Chirikov1984,Meiss1986}. Such examples occur in many physically realistic systems studied in the current literature (see e.g. \cite{Skokos2008,Flach2009,Johansson2009,Skokos2009}).

In this paper, we plan to focus on such ``weakly chaotic'' regimes of the first and last normal modes of the Fermi-Pasta-Ulam (FPU)-$\alpha$ 1-dimensional particle chain under fixed boundary conditions. We thus demonstrate, by means of numerical experiments according to the CLT, that pdfs of sums of orbit components at low energies \textit{do not} rapidly converge to a Gaussian, but are well approximated for long integration times by the so-called $q$-Gaussian function \cite{Tsallisbook2009}
\begin{equation}\label{q_gaussian}
P(s)=d\exp_q({-\beta s^2})\equiv d\biggl[1-(1-q)\beta s^{2}\biggr]^{\frac{1}{1-q}}
\end{equation}
where the index satisfies $1<q<3$, $\beta$ is an arbitrary parameter and $d$ is a normalization constant. At longer times, of course, weakly chaotic orbits are expected to eventually escape to larger chaotic seas, where obstruction by islands and cantori is less dominant and the dynamics is more uniformly ergodic. This transition is associated with energy equipartition among all linear modes and is characterized by $q$-Gaussian pdfs \eqref{q_gaussian} whose index decreases to $q=1$, representing the limit at which the pdf becomes a Gaussian.

Thus, in our models, $q$-Gaussian distributions represent \textit{quasi-stationary states} (QSS) that are often very \textit{long-lived}, especially inside ``thin'' chaotic layers, where the Lyapunov exponents are small. This suggests that it might be useful to study these pdfs (and their associated $q$ values) for sufficiently long times to derive information about the long term dynamics of the QSS and their connection to energy equipartition among all degrees of freedom. 

In what follows, after introducing our methodology in Sec. \ref{CLT_approach}, we begin in Sec. \ref{FPU_a_model_section} by describing the system under study and some of its main properties. Next, in Sec. \ref{single_site_excitation_FPU_a_model_section} we study two special solutions of the Fermi-Pasta-Ulam-$\alpha$ (FPU-$\alpha$) 1-dimensional chain of $N=31$ particles under \textit{fixed} boundary conditions. These solutions correspond to: (a) The excitation of the $k=1$ linear mode, which has been related to the FPU paradox \cite{Fermi1955} and (b) the excitation of the last ($k=31$) linear mode, at energies where the nonlinear continuations ($\alpha>0$) of these modes have turned unstable. In case (a), we find that chaotic diffusion phenomena (related to the breakdown of the so-called FPU recurrences) are characterized by $q$-Gaussian pdfs with $q$ decreasing to 1 as energy increases and equipartition sets in at earlier and earlier times. On the other hand, case (b) shows persistent quasi-periodic behavior until, it suddenly becomes strongly chaotic and equipartition abruptly occurs. These results may be explained by examining the exponential localization properties of the averaged modal energies $\bar{E}_k$ of the two solutions in $k$-space.

%%%%%%%%%%%%%%%%%%%%%%%%%%%%%%%%%%%%%%%%%%%%%%%%%%%%%%%%%%%%%%%%%%%%%%%%%%%%
%%%%%%%%%%%%%%%%%%%%%%%%%%%%%%%%%%%%%%%%%%%%%%%%%%%%%%%%%%%%%%%%%%%%%%%%%%%%
\section{Statistical distributions of chaotic QSS and their computation}\label{CLT_approach}
%%%%%%%%%%%%%%%%%%%%%%%%%%%%%%%%%%%%%%%%%%%%%%%%%%%%%%%%%%%%%%%%%%%%%%%%%%%%
%%%%%%%%%%%%%%%%%%%%%%%%%%%%%%%%%%%%%%%%%%%%%%%%%%%%%%%%%%%%%%%%%%%%%%%%%%%%

The problem we investigate is described by an autonomous $N$ degree of freedom Hamiltonian function of the form
\begin{equation}
H\equiv H(q(t),p(t))=H(q_1(t),\ldots,q_N(t),p_1(t),\ldots,p_N(t))=E\label{Ham_fun}
\end{equation}
where $(q_k(t),p_k(t))$ are respectively the positions and momenta representing the solution in phase space at time $t$. As is well-known, these solutions can be periodic, quasi-periodic or chaotic depending on the initial condition $(q(0)$,$p(0))$ and the value of the total energy $E$. What we wish to study here is the statistics of such systems in regimes of ``weakly'' chaotic motion, where the Lyapunov exponents \cite{Benettin1980A,Benettin1980B,Eckmann1985,Skokos2010} are positive but very small. Such situations often arise when one considers solutions which diffuse slowly in thin chaotic layers and wander through a complicated network of higher order resonances, often sticking for very long times to the boundaries of islands constituting the so-called ``edge of chaos'' regime \cite{Aizawa1984,Chirikov1984,Meiss1986,Tsallisbook2009}.

At this point one can ask the following questions: How long do these ``weakly'' chaotic states last? Assuming they are quasi-stationary, can we describe them \textit{statistically} by following some of their chaotic orbits? What type of distributions characterize these QSS and how could one connect them to the actual dynamics of the solutions of the corresponding Hamiltonian system? 

Following an approach inspired by the well-known Central Limit Theorem \cite{Rice1995}, we shall use the solutions of Hamilton's equations of motion
\begin{equation}\label{eq:Hamiltonian_ODEs}
\frac{{dq}_{k}}{dt}=\frac{\partial H}{{\partial
p}_{k}},\qquad\frac{{dp}_{k}}{dt}=-\frac{\partial H}{{\partial
q}_{k}},\;k=1,\ldots,N
\end{equation}
to construct \textit{distributions} of suitably rescaled sums of $M$ values of a generic observable $\eta_i=\eta(t_i)\;(i=1,\ldots,M)$ which depends linearly on the components of the solution. If these are viewed as independent and identically distributed random variables (in the limit $M\rightarrow\infty$), we may evaluate their sum
\begin{equation}\label{sums_CLT}
S_M^{(j)}=\sum_{i=1}^M\eta_i^{(j)}
\end{equation}
for $j=1,\ldots,N_{ic}$ different initial conditions. Thus, we can study the statistical behavior of the variables $S_M^{(j)}$, centered about their mean value $\langle S_M^{(j)}\rangle=\frac{1}{N_{ic}}\sum_{j=1}^{N_{ic}}\sum_{i=1}^{M}\eta_i^{(j)}$ and rescaled by their standard deviation $\sigma_M$
\begin{equation}
s_M^{(j)}\equiv\frac{1}{\sigma_M}\Bigl(S_M^{(j)}-\langle S_M^{(j)}\rangle \Bigr)=\frac{1}{\sigma_M}\Biggl(\sum_{i=1}^M\eta_i^{(j)}-\frac{1}{N_{ic}}\sum_{j=1}^{N_{ic}}\sum_{i=1}^{M}\eta_i^{(j)}\Biggl)
\end{equation}
where
\begin{equation}
\sigma_M^2=\frac{1}{N_{ic}}\sum_{j=1}^{N_{ic}}\Bigl(S_M^{(j)}-\langle S_M^{(j)}\rangle \Bigr)^2=\langle S_M^{(j)2}\rangle -\langle S_M^{(j)}\rangle^2.
\end{equation}
Plotting the normalized histogram of the probabilities $P(s_M^{(j)})$ as a function of $s_M^{(j)}$, we then compare our pdfs with a $q$-Gaussian function of the form
\begin{equation}\label{q_gaussian_distrib}
P(s_M^{(j)})=d\exp_q({-\beta s_M^{(j)2}})\equiv d\biggl[1-(1-q)\beta s_M^{(j)2}\biggr]^{\frac{1}{1-q}}
\end{equation}
where $q$ is the so-called entropic index, $\beta$ is a free parameter and $d$ a normalization constant \cite{Tsallisbook2009}. Note that, in the limit $q\rightarrow 1$, \eqref{q_gaussian_distrib} tends to the well-known Gaussian distribution, i.e. $\exp_q(-\beta x^2)\rightarrow\exp(-\beta x^2)$.

It can be shown that the $q$-Gaussian distribution \eqref{q_gaussian_distrib} is normalized when
\begin{equation}\label{beta-$q$-Gaussian}
\beta=d\sqrt{\pi}\frac{\Gamma\Bigl(\frac{3-q}{2(q-1)}\Bigr)}{(q-1)^{\frac{1}{2}}\Gamma\Bigl(\frac{1}{q-1}\Bigr)}
\end{equation}
where $\Gamma$ is the Euler $\Gamma$ function \cite{Tsallisbook2009}. Clearly, \eqref{beta-$q$-Gaussian} shows that the allowed values of $q$ are $1<q<3$.

The index $q$ appearing in \eqref{q_gaussian_distrib} is connected with the Tsallis entropy \cite{Tsallisbook2009}
\begin{equation}\label{Tsallis entropy}
S_q=k\frac{1-\sum_{i=1}^W p_i^q}{q-1}\mbox{ with }\sum_{i=1}^W p_i=1
\end{equation}
where $i=1,\ldots,W$ counts the microstates of the system, each occurring with a probability $p_i$ and $k$ is the so-called Boltzmann universal constant. Just as the Gaussian distribution represents an extremal of the BG entropy $S_{\mbox{BG}}\equiv S_1=k\sum_{i=1}^W p_i\ln p_i$, so is the $q$-Gaussian \eqref{q_gaussian_distrib} derived by optimizing the Tsallis entropy \eqref{Tsallis entropy} under appropriate constraints. Systems characterized by the Tsallis entropy are said to lie at the ``edge of chaos'' and are significantly different than BG systems, in the sense that their entropy is non-additive and generally non-extensive \cite{Tsallisbook2009,TsallisTirnakli2010}. 

We now describe the numerical approach we follow to calculate the above pdfs. First, we specify an observable denoted by $\eta(t)$ as one (or a linear combination) of the components of the position vector $q(t)$ of a chaotic solution of \eqref{eq:Hamiltonian_ODEs}, located initially at $(q(0),p(0))$.
Assuming that the orbit visits all parts of a QSS during the integration interval $0\leq t\leq t_f$, we divide $t_f$ into $N_{ic}$ equally spaced, consecutive time windows, which are long enough to contain a significant part of the orbit. Next, we subdivide each such window into a number $M$ of equally spaced subintervals and calculate the sum $S_M^{(j)}$ of the values of the observable $\eta(t)$ at the \textit{right edges} of these subintervals (see eq. \eqref{sums_CLT}).

In this way, we treat the beginning of every time window as a new initial condition and repeat this process $N_{ic}$ times, to obtain as many sums as required for reliable statistics. Consequently, at the end of the integration, we compute the average and standard deviation of the sums \eqref{sums_CLT}, evaluate the $N_{ic}$ rescaled quantities $s_M^{(j)}$ and plot the histogram $P(s_M^{(j)})$ of their distribution.

As we shall see in the next sections, in regions of ``weak chaos'' these distributions are well-fitted by a $q$-Gaussian of the form \eqref{q_gaussian_distrib} for fairly long time intervals. However, for longer times (or higher energies), as the orbit diffuses in domains of ``strong chaos'', a Gaussian pdf (with $q\simeq1$) is expected to describe the dynamics.

%%%%%%%%%%%%%%%%%%%%%%%%%%%%%%%%%%%%%%%%%%%%%%%%%%%%%%%%%%%%%%%%%%%%%%%%%%%%
%%%%%%%%%%%%%%%%%%%%%%%%%%%%%%%%%%%%%%%%%%%%%%%%%%%%%%%%%%%%%%%%%%%%%%%%%%%%
\section{The Fermi-Pasta-Ulam-$\alpha$ model}\label{FPU_a_model_section}
%%%%%%%%%%%%%%%%%%%%%%%%%%%%%%%%%%%%%%%%%%%%%%%%%%%%%%%%%%%%%%%%%%%%%%%%%%%%
%%%%%%%%%%%%%%%%%%%%%%%%%%%%%%%%%%%%%%%%%%%%%%%%%%%%%%%%%%%%%%%%%%%%%%%%%%%%

The well-known FPU model is a Hamiltonian lattice of $N$ particles (degrees of freedom) with a potential that depends only on nearest neighbor particle interactions. Its name was given after the pioneering work by E. Fermi, J. Pasta, S. Ulam and M. Tsingou in the 1950's \cite{Fermi1955}, which demonstrated the lack of thermalization of the system in contrast to what is expected by BG Statistical Mechanics. In particular, they reported the absence of energy transfer to all normal modes of the system even after very long integration times, due to the so-called FPU recurrences, where the energy returns from neighboring normal modes to the first one that was initially excited. This phenomenon, termed ``the FPU paradox'', still remains an open problem, although much progress has been made to date towards understanding its dynamical behavior in the thermodynamic limit of $E$ and $N$ becoming arbitrarily large, with $E/N=$const.

More specifically, in the present work we deal with the FPU-$\alpha$ model of $N$ identical particles on a 1-dimensional chain described by the Hamiltonian
\begin{eqnarray}\label{fpuham}
H= {\frac{1}{2}}\sum_{n=1}^{N} p_n^2 + {\frac{1}{2}}\sum_{n=0}^{N}(q_{n+1}-q_n)^2 + {\frac{\alpha}{3}}\sum_{n=0}^{N}(q_{n+1}-q_n)^3=E
\end{eqnarray}
where $q_{n}$ is the displacement of the $n$th particle from its equilibrium position, $p_{n}$ is the corresponding momentum, $\alpha\geq0$ is a real constant and $E$ is the fixed energy of the system. Moreover, we shall consider here only fixed boundary conditions, i.e. $q_0=q_{N+1}=0$ following \cite{Fermi1955}.

As is already known, under the canonical normal mode transformation
\begin{eqnarray}
q_n &=&\sqrt{\frac{2}{N+1}}\sum_{k=1}^{N} Q_k\sin\left({kn\frac{\pi}{N+1}}\right),\label{lintra1}\\
 p_n&=&\sqrt{\frac{2}{N+1}}\sum_{k=1}^{N}
P_k\sin\left({kn\frac{\pi}{N+1}}\right),\;n=1,\ldots,N,\label{lintra2}
\end{eqnarray}
system (\ref{fpuham}) can be studied following the time evolution of the harmonic normal modes of the linear case (i.e. $\alpha =0$), whose energies $E_k$ are given by
\begin{equation}\label{harmonicene}
E_k=P_k^2+ \omega_k ^2 Q_k^2
\end{equation}
where
\begin{equation}\label{fpuspec}
\omega_k=2\sin\left({\frac{k \pi}{2(N+1)}}\right),\;k=1,\ldots,N
\end{equation}
are the harmonic frequencies. Taking the time averaged harmonic spectra
\begin{equation}\label{harmonicaveene}
\bar{E} _k (t)= \frac{1}{t} \int _0 ^t E_k(s) ds
\end{equation}
one expects to find that the system reaches equipartition of its total energy among all its normal modes if $\bar{E}_k \rightarrow E/N\;\forall k$ as $t\rightarrow \infty$. 

%%%%%%%%%%%%%%%%%%%%%%%%%%%%%%%%%%%%%%%%%%%%%%%%%%%%%%%%%%%%%%%%%%%%%%%%%%%%
%%%%%%%%%%%%%%%%%%%%%%%%%%%%%%%%%%%%%%%%%%%%%%%%%%%%%%%%%%%%%%%%%%%%%%%%%%%%
\section{Single site excitations in the FPU-$\alpha$ model}\label{single_site_excitation_FPU_a_model_section}
%%%%%%%%%%%%%%%%%%%%%%%%%%%%%%%%%%%%%%%%%%%%%%%%%%%%%%%%%%%%%%%%%%%%%%%%%%%%
%%%%%%%%%%%%%%%%%%%%%%%%%%%%%%%%%%%%%%%%%%%%%%%%%%%%%%%%%%%%%%%%%%%%%%%%%%%%

We now proceed to study two different cases of single site excitations in the FPU-$\alpha$ system (see eq. \eqref{fpuham}) to investigate how energy spreads between all normal modes. The first is the well-known case of the excitation of the first normal mode $k=1$, which is the one studied originally by Fermi, Pasta and Ulam in \cite{Fermi1955} and the second one concerns the excitation of the last mode $k=N$, which correspond to the longest and shortest wavelength of the spectrum respectively. As we discuss below, they turn out to be strikingly different in the way they lead to the excitation of the full spectrum of $E_k,\;k=1,\ldots,N$.

For the numerical integration of our trajectories, we typically use throughout the paper Yoshida's fourth order symplectic integrator \cite{Yoshida1990} with a fixed time step varying between $0.01 - 0.05$ depending on the particular energy $E$ of Hamiltonian \eqref{fpuham}, which always results in a relative energy error of the order of $10^{-6} - 10^{-7}$. For the computation of the Lyapunov exponents and the Generalized Alignment Index (GALI) \cite{Skokos2007} using the above symplectic integrator (or any symplectic integrator in general), we apply the methodology of the {\em Tangent Dynamics Hamiltonian} proposed in \cite{Skokosetal2010} which is suitable for the evolution of deviation vectors in the tangent space of the orbit under study. Having thus access to the deviation vectors, we compute the Lyapunov exponents following the standard method of \cite{Benettin1980A,Benettin1980B} and GALI$_i,\;i\geq2$ and use a numerically faster implementation of the GALIs which is suitable especially for multi-dimensional systems, called the ``Linear Dependence Index'' (LDI) as in \cite{Antonopoulosetal2006}.

Following \cite{Dresden}, we study the case of $N=31$ particles of the FPU-$\alpha$ system \eqref{fpuham} with $\alpha =0.33$. In particular, in \cite{Dresden} a detailed study of energy diffusion is carried out, from the first normal mode $k=1$ which is initially excited, to all remaining modes of the spectrum and an estimate of the time needed to reach equipartition is presented. Here, we are interested in focusing on regimes closer to the so-called ``edge of chaos'', i.e. phase space regions where diffusion processes are extremely weak due to the ``stickiness'' phenomenon \cite{Aizawa1984,Chirikov1984,Meiss1986,Skokos2008}. Aided by the GALI method and the computation of Lyapunov exponents, we use concepts of Non-Extensive Statistical Mechanics \cite{Tsallisbook2009} to show that $q$-Gaussian distributions can describe quite satisfactorily weakly chaotic dynamics in excitations of the $k=1$ mode, while they identify a distinctly different sharp transition to strong chaos and equipartition in excitations of the $k=31$ mode. Our results are in agreement with similar investigations carried out in \cite{Antonopoulosetal2010}, concerning QSS in multi-dimensional FPU-$\beta$ models.

%%%%%%%%%%%%%%%%%%%%%%%%%%%%%%%%%%%%%%%%%%%%%%%%%%%%%%%%%%%%%%%%%%%%%%%%%%%%
%%%%%%%%%%%%%%%%%%%%%%%%%%%%%%%%%%%%%%%%%%%%%%%%%%%%%%%%%%%%%%%%%%%%%%%%%%%%
\subsection{Excitations of the first normal mode}
%%%%%%%%%%%%%%%%%%%%%%%%%%%%%%%%%%%%%%%%%%%%%%%%%%%%%%%%%%%%%%%%%%%%%%%%%%%%
%%%%%%%%%%%%%%%%%%%%%%%%%%%%%%%%%%%%%%%%%%%%%%%%%%%%%%%%%%%%%%%%%%%%%%%%%%%%

\begin{figure}
\centering
\includegraphics[scale=0.3 ]{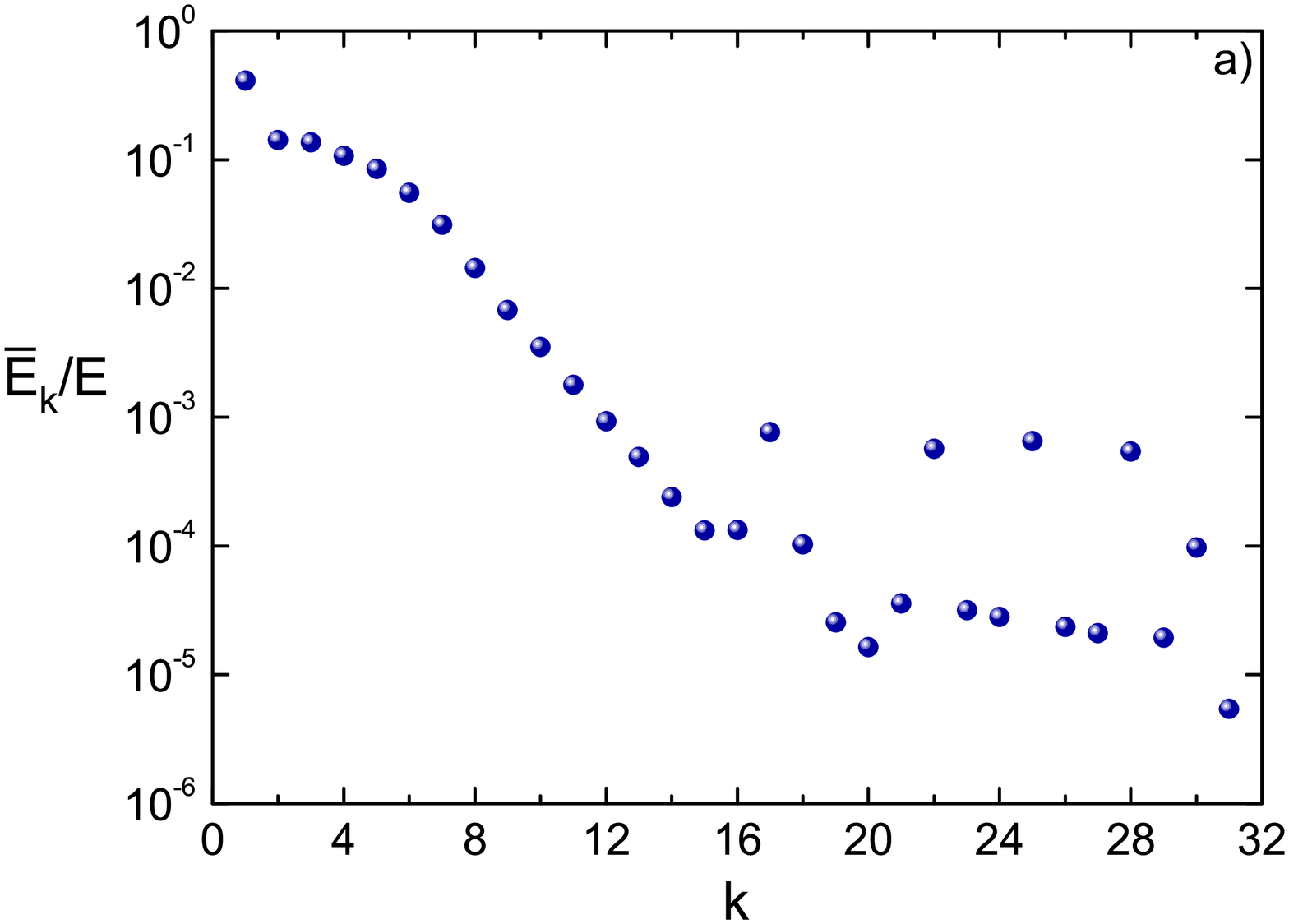}% Here is how to import EPS art
\includegraphics[scale=0.3 ]{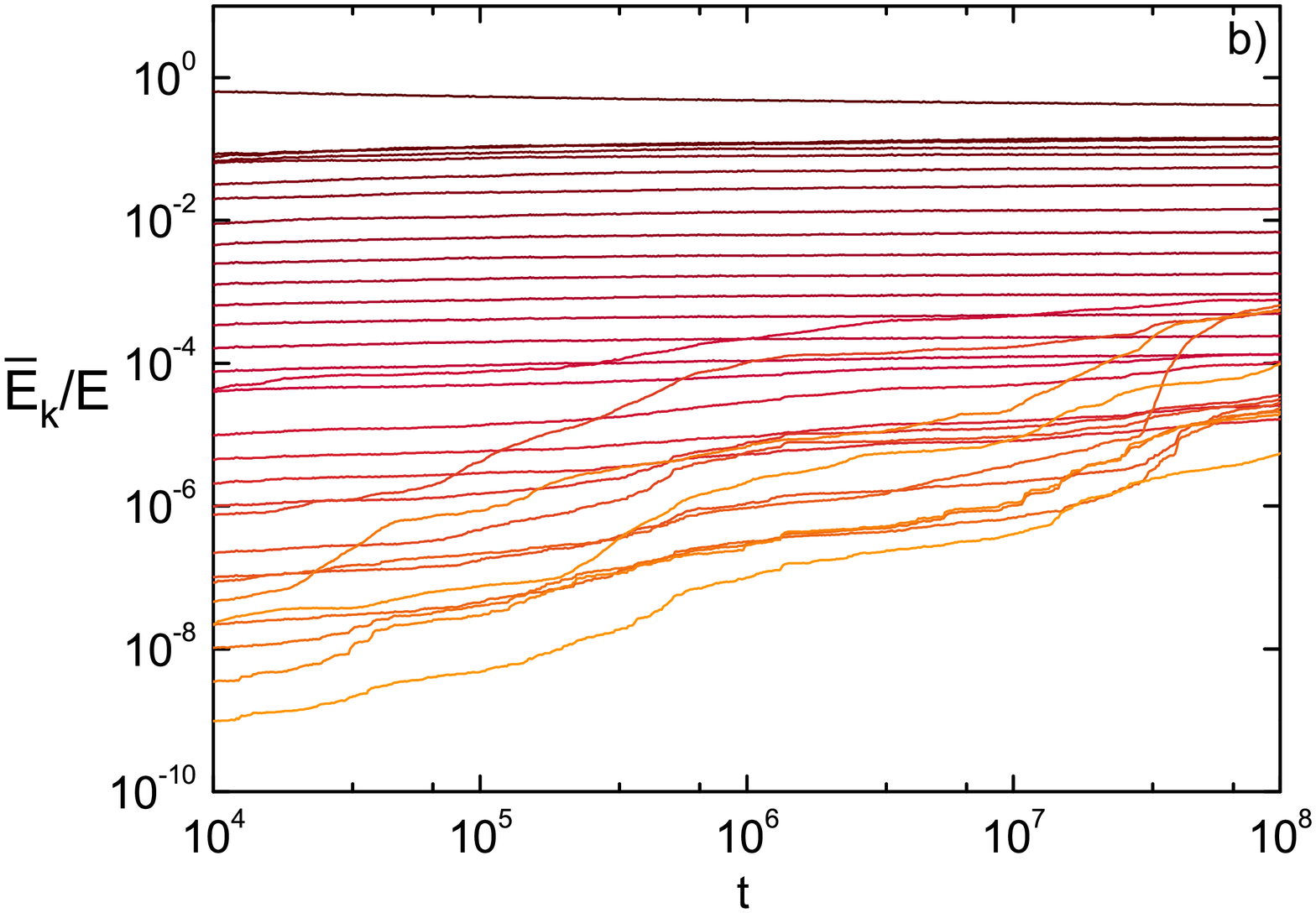}\\% Here is how to import EPS art
\includegraphics[scale=0.3]{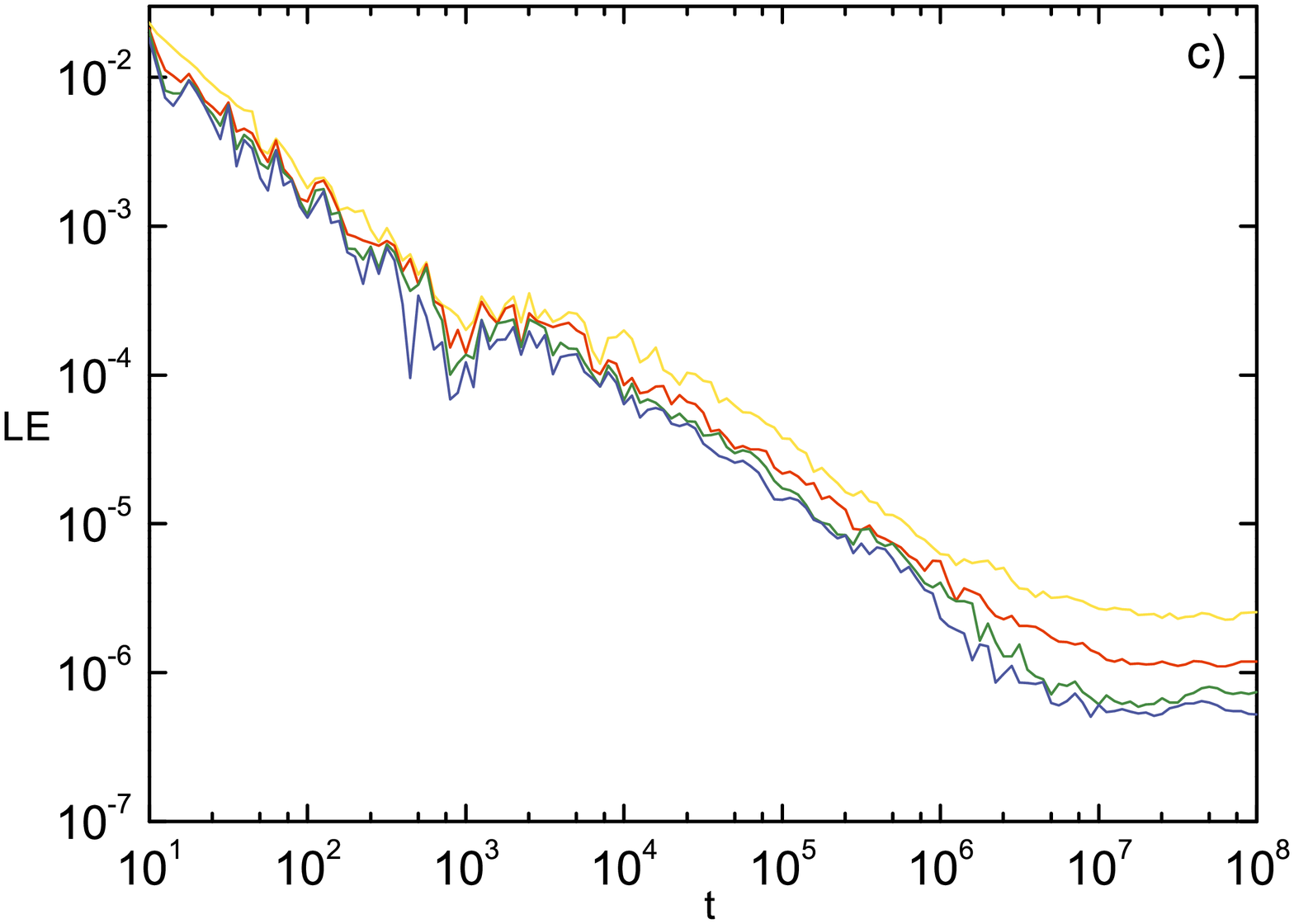}
\includegraphics[scale=0.3 ]{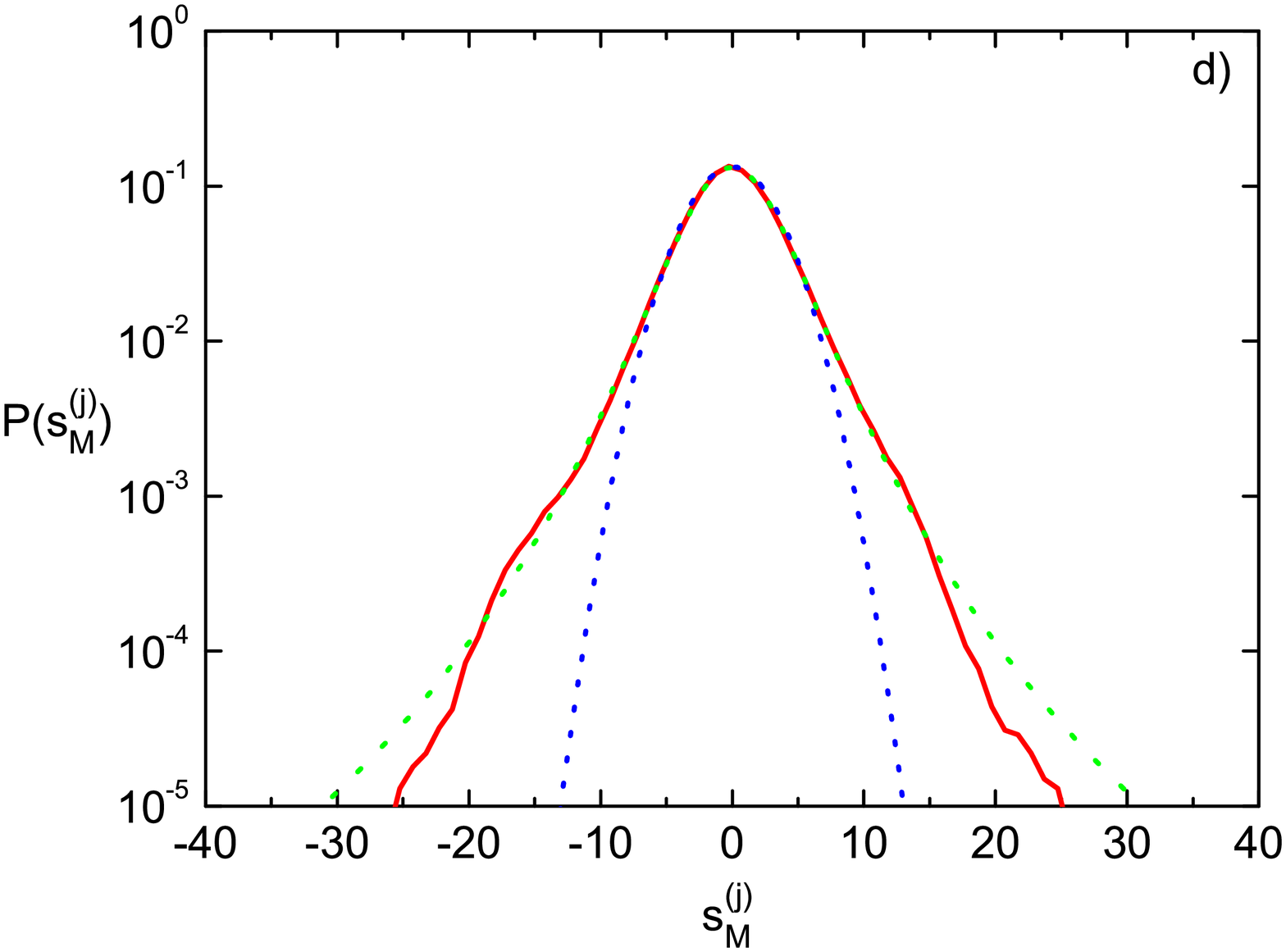}
\caption{FPU-$\alpha$ system with $N=31$, $\alpha =0.33$ and $E=0.19$. The initial condition is such that all the energy $E$ is given to the first normal mode $k=1$ (see eq. \eqref{lintra1} and \eqref{lintra2}). In panel a) we plot the normalized averaged energy spectrum at $t_f=10^8$, while panel b) shows the time evolution of the averaged energy spectra $\bar{E}_k/E$ of eq. \eqref{harmonicaveene}. Panel c) presents the time evolution of the four biggest Lyapunov exponents. Panel d) is a plot (in lin - log scale) of the numerical data (red curve), $q$-Gaussian (green curve) and Gaussian (blue curve) distributions that correspond to final integration time $t_f=10^{11}$ using $N_{ic}=10^6$ time windows and $M=100$ terms in the computation of the sums for the observable $\eta=q_1+q_2$. In this case the numerical fitting with a $q$-Gaussian gives $q\approx1.342$ with $\chi^2\approx2.41\times10^{-5}$.}\label{fig:q=1_E=0.19}
\end{figure}

Let us start our study exciting the first normal mode only, i.e. $k=1$ in eqs. \eqref{lintra1} and \eqref{lintra2}. In Fig.~\ref{fig:q=1_E=0.19}a) we present in logarithmic scale the averaged energy spectra of the system at time $t=10^8$ for the energy $E=0.19$. As has already been pointed out in \cite{PB,BP,Dresden}, the system is in a metastable state, where few (low $k$) modes share the total energy of the system, forming a ``natural packet'' \cite{Gio1}, while all the rest gain a small amount of it and remain for long times exponentially localized in normal mode space. Although, for these parameters and initial conditions, the first normal mode seems to dominate the motion, a rather weak diffusion takes place in the highest modes of the spectrum, as reported and quantified in \cite{Dresden}. In Fig.~\ref{fig:q=1_E=0.19}b) the evolution of the averaged harmonic energies $\bar{E}_k/E$ of eq. \eqref{harmonicene} with $N=31$, $\alpha =0.33$ and $E=0.19$ is presented in log - log scale. The lower modes are depicted by the darkest colors and the higher by the lighter ones. As is evident, a weak diffusion of energy from the lowest to the highest modes is revealed, resulting in a slow increase of the $E_k$ values for large $k$. This suggests that we can expect energy equipartition after a long time interval $t_f\approx10^7$ during which the maximal Lyapunov exponents decrease towards zero, saturating thereafter at values close to $10^{-6}$, as shown on panel c) of Fig. \ref{fig:q=1_E=0.19}.

It is therefore important to study the behavior in the immediate neighborhood of this mode at $E=0.19$, in terms of statistical distributions by the methods described in Sec. \ref{CLT_approach}. In panel d) of Fig. \ref{fig:q=1_E=0.19} we plot the numerical (red curve), $q$-Gaussian (green curve) and Gaussian (blue curve) distributions that correspond to final integration time $t_f=10^{11}$ using $N_{ic}=10^6$ time windows and $M=100$ terms in the computation of the sums for $\eta=q_1+q_2$. Our results show that this weakly diffusive motion towards equipartition is described by a pdf that is well-fitted by a $q$-Gaussian of the form \eqref{q_gaussian_distrib}, with $q\approx1.416,\;\chi^2\approx5\times10^{-4}$ for $t_f=10^{9}$ and $q\approx1.342,\;\chi^2\approx2.41\times10^{-5}$ for $t_f=10^{11}$ (see Fig. \ref{fig:q=1_E=0.19}d)).

%%%%%%%%%%%%%%%%%%%%%%%%%%%%%%%%%%%%%%%%%%%%%%%%%%%%%%%%%%%%%%%%%%%%%%%%%%%%
%%%%%%%%%%%%%%%%%%%%%%%%%%%%%%%%%%%%%%%%%%%%%%%%%%%%%%%%%%%%%%%%%%%%%%%%%%%%
\subsection{Excitations of the last normal mode}\label{subsection_last_mode}
%%%%%%%%%%%%%%%%%%%%%%%%%%%%%%%%%%%%%%%%%%%%%%%%%%%%%%%%%%%%%%%%%%%%%%%%%%%%
%%%%%%%%%%%%%%%%%%%%%%%%%%%%%%%%%%%%%%%%%%%%%%%%%%%%%%%%%%%%%%%%%%%%%%%%%%%%

In system \eqref{fpuham} of $N=31$ degrees of freedom, the last normal mode is that of $k=31$ and has the shortest wave length. Since its frequency $\omega_{31}\simeq 1.9975\simeq 2$ lies outside the acoustic resonance regime, $k=31$ is more strongly localized in modal space than the $k=1$ mode and thus it is interesting to study it in its own right.
\begin{figure}
\centering
\includegraphics[scale=0.3]{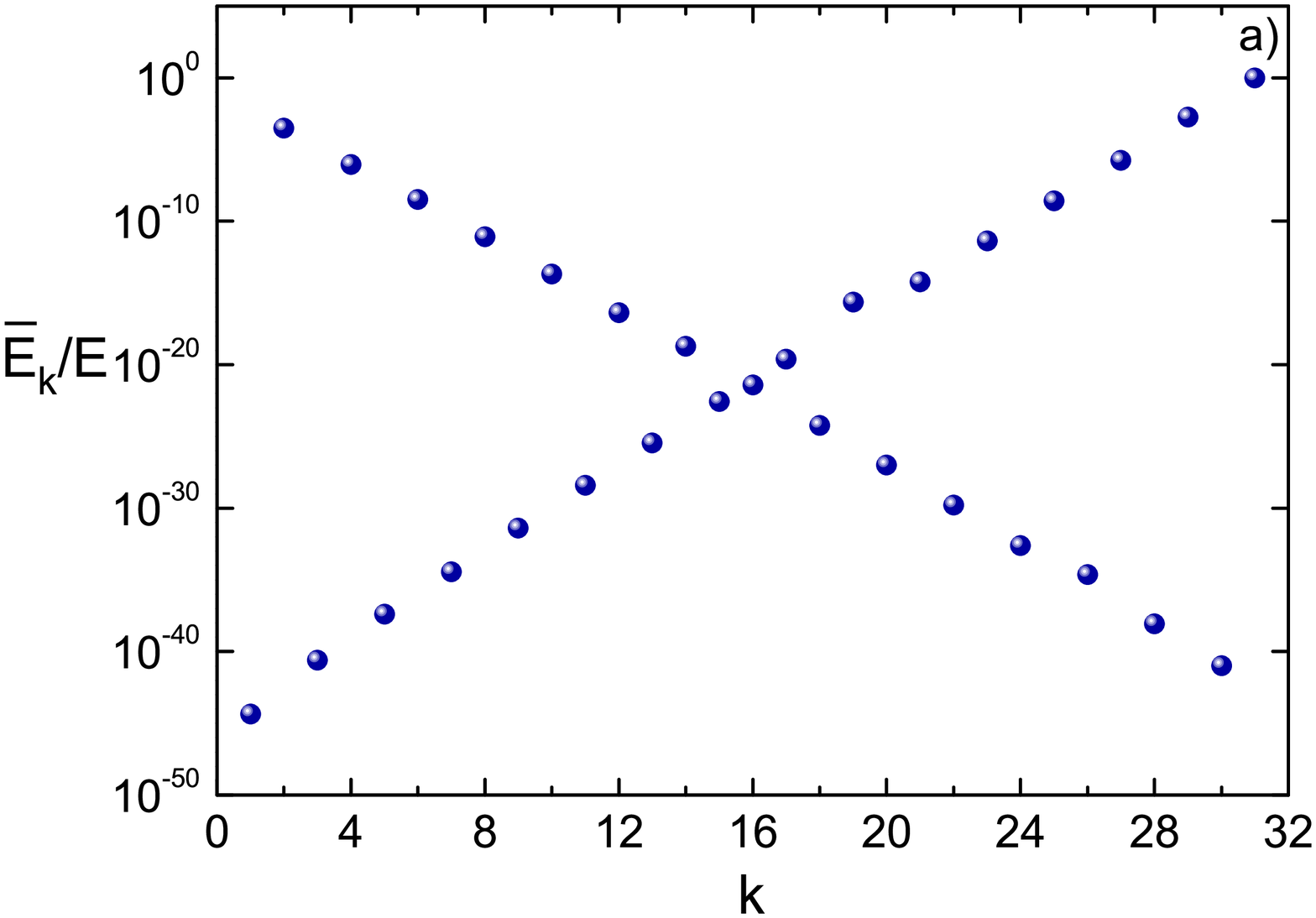}% Here is how to import EPS art
\includegraphics[scale=0.3]{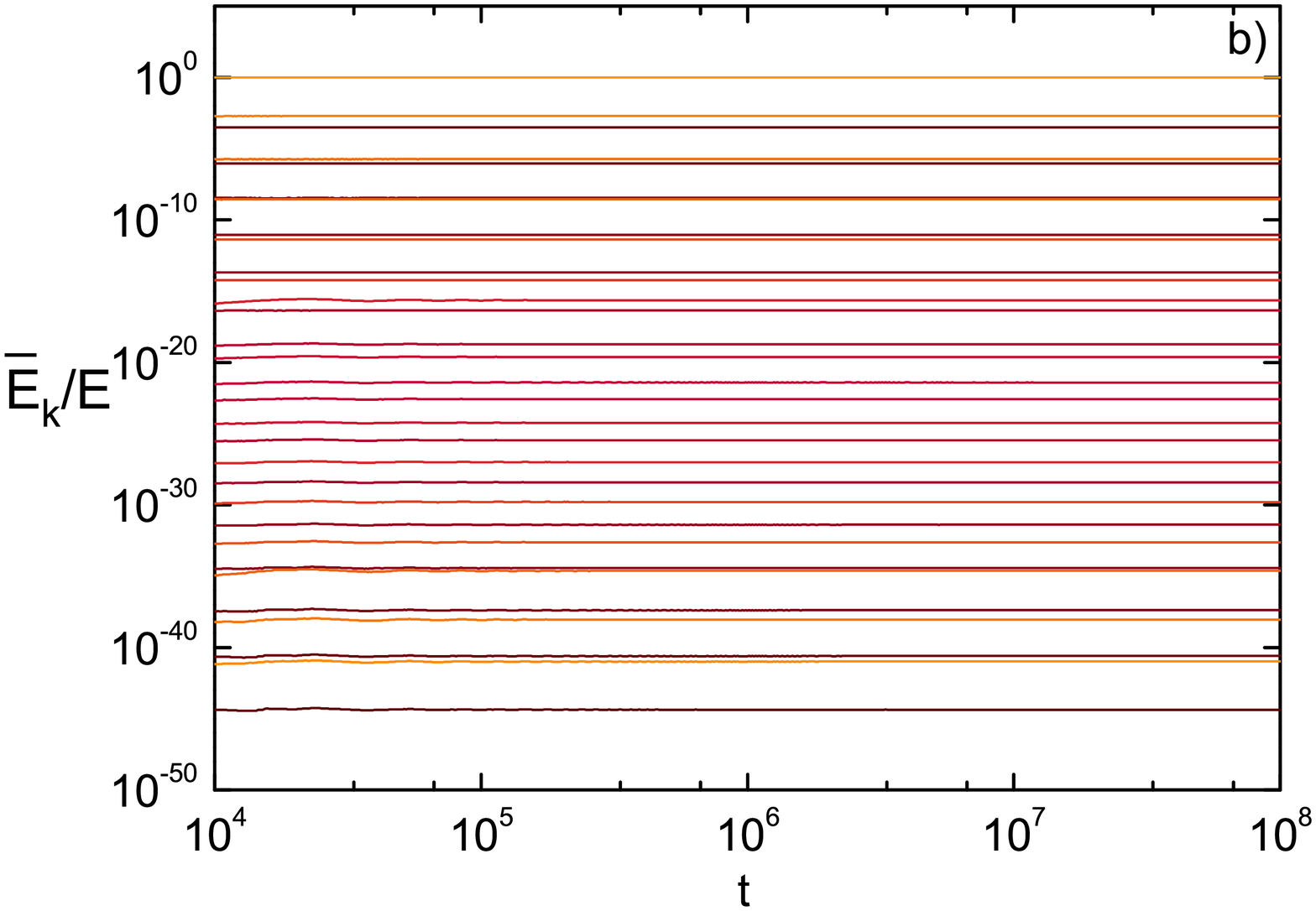}\\% Here is how to import EPS art
\includegraphics[scale=0.3 ]{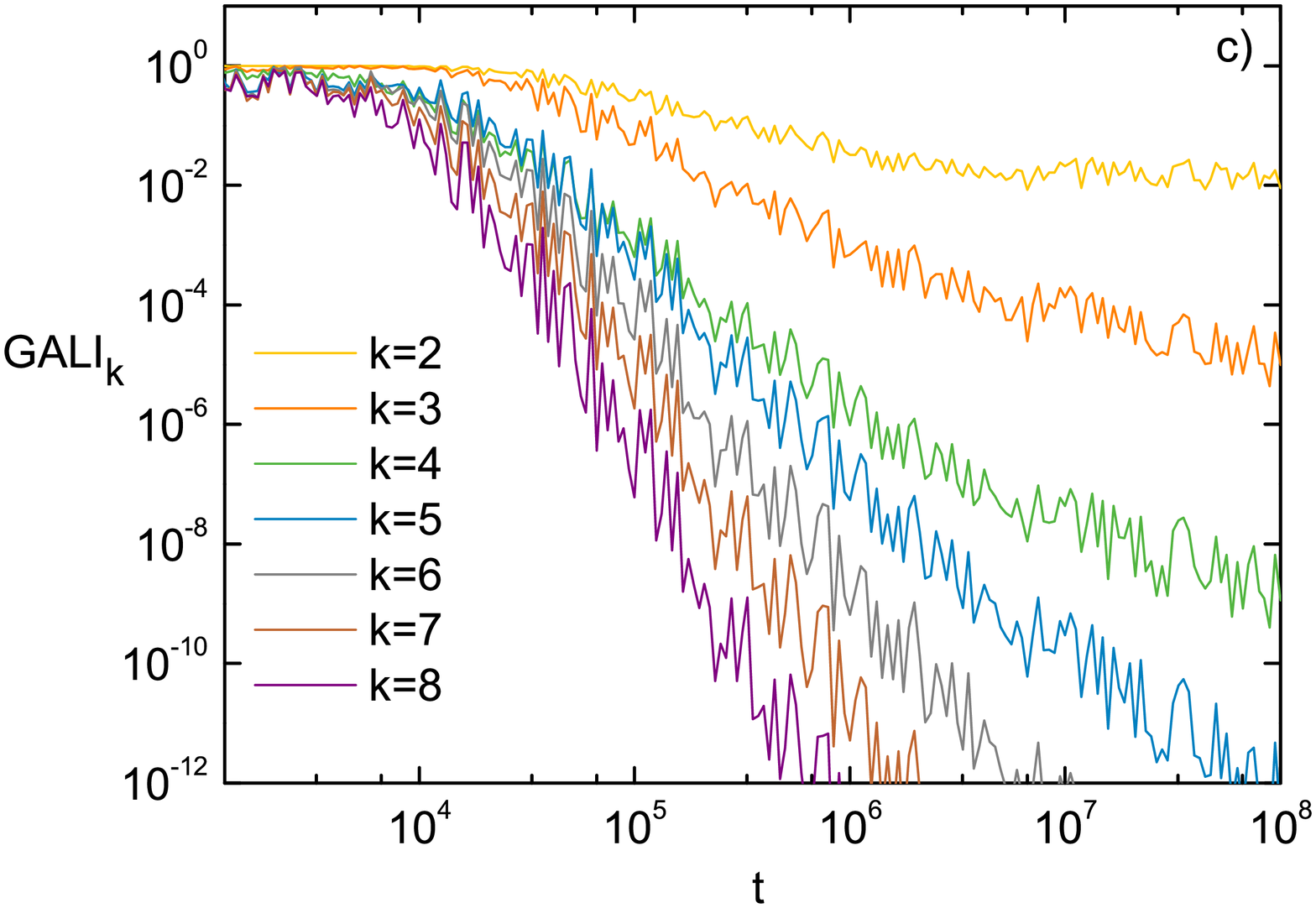}
\caption{FPU-$\alpha$ system with $N=31$, $\alpha =0.33$ and $E=0.19$. The initial condition here is such that all the energy $E$ is attributed to the last normal mode $k=31$. Panel a) shows a plot of the normalized averaged energy spectrum at $t_f=10^8$, exhibiting perfect exponential localization. In panel b) the time evolution of the averaged energy spectra $\bar{E}_k/E$ of eq. \eqref{harmonicaveene} shows complete absence of diffusion, while in panel c) the time evolution of the LDI$_i=$GALI$_i,\;i=2,\ldots,8$ indicates that the motion is indeed quasi-periodic lying on a 2-dimensional torus.}\label{fig:q=31_E=0.19}
\end{figure}

In Fig.~\ref{fig:q=31_E=0.19}a) we present the averaged energy spectra of the $k=31$ mode at $t=10^8$ and in panel b) the time evolution of its averaged energies $\bar{E}_k/E$ for $E=0.19$ and $\alpha =0.33$. In \cite{Flach1,Flach2,Flach3,Flach4} the authors have predicted that this kind of trajectories, for values of the parameters as considered here, lie close to a q-breather and thus the energy spectra that appear in Fig. \ref{fig:q=31_E=0.19}a) are approximately the same as those of the q-breather. The energy spectrum of the latter arises from the continuation of the simple periodic orbit that corresponds to mode $k=31$ of the linear system (see eq. \eqref{fpuham} for $\alpha =0$). Our results also demonstrate that the FPU trajectory arising from the excitation of the mode $k=31$ is close to a q-breather, since the hierarchy of the harmonic energies $E_k$ of the q-breather, calculated by using the proposition in \cite{PenatiFlach} for a single mode excitation or in \cite{Christodoulidi2010}, \cite{Christodoulidi2011} for multiple mode excitation, is almost the same as the trajectory shown in Fig.~\ref{fig:q=31_E=0.19}a). 

In particular, in \cite{Christodoulidi2011} it is predicted, using the Poincar\'{e} - Lindstedt method, that if we start with $k=31$ at zeroth order, the modes $k_1=2$, $k_2=N-2=29$, $k_3=4$, $k_4=N-4=27$ etc. will be successively excited at higher orders (for more details see the Appendix below). Knowing the sequence they follow, $k_1,k_2,\ldots,k_N$, we can identify these modes with those called ``tail modes'' in the literature \cite{PenatiFlach,Dresden} as having the lowest energy values. Thus, choosing these modes to be $k_{N/2},\ldots,k_{N}$, one expects that the energy will diffuse from the strongly localized and dominant modes $k_{1},\ldots,k_{N/2-1}$ to the tail modes. However, as we deduce from Fig. \ref{fig:q=31_E=0.19}b), there is no presence of energy diffusion to the tail modes of the system at least up to $t_f=10^8$. 

This suggests that the localization phenomenon is very strong compared with that of the $k=1$ case, on the same energy manifold $E=0.19$. By computing the LDI$_i=$GALI$_i,\;i=2,\ldots,8$ in Fig. \ref{fig:q=31_E=0.19}c), we assert that the motion remains ordered at least up to $t_f=10^8$ that we have checked numerically, in fully agreement with what we have found in panel b) of the same figure. The same finding is also justified by the fact that the four maximal Lyapunov exponents (not shown here) keep decreasing to zero up to the same final integration time. We conclude, therefore, that exciting only $k=31$ yields a regular, quasi-periodic orbit, at least up to $t_f=10^8$, apparently lying on a 2-dimensional torus, since GALI$_2$ is the only index that remains nearly constant in this time interval.

This striking difference in the time evolution of the harmonic energies in the above two cases is due to the acoustic resonance of $k=1$ with a few nearby modes, i.e. $\omega _k \simeq k\;\omega _1$. These modes are known to compose a ``natural packet'' and undergo an internal equipartition of the total energy. This natural packet has width equal to $\mu =\alpha ^{1/2} \Bigr({\frac{E}{N+1}}\Bigl)^{1/4} (N+1)$, as predicted in \cite{Gio1,Gio2,BP,PB}. For example, in Fig.~\ref{fig:q=1_E=0.19} we obtain $\mu=5.10279$.

\begin{figure}
\centering
\includegraphics[scale=0.3 ]{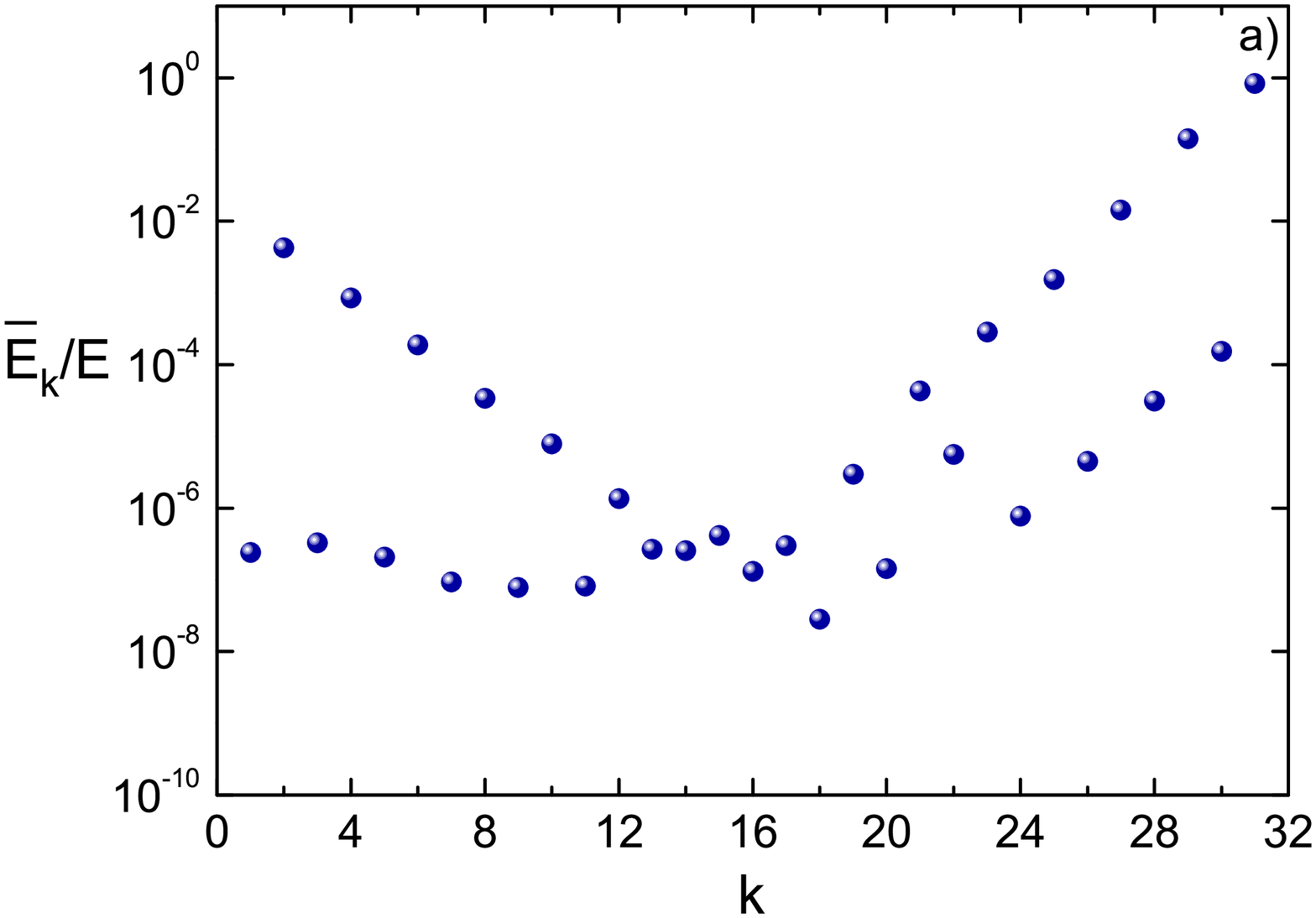}% Here is how to import EPS art
\includegraphics[scale=0.3 ]{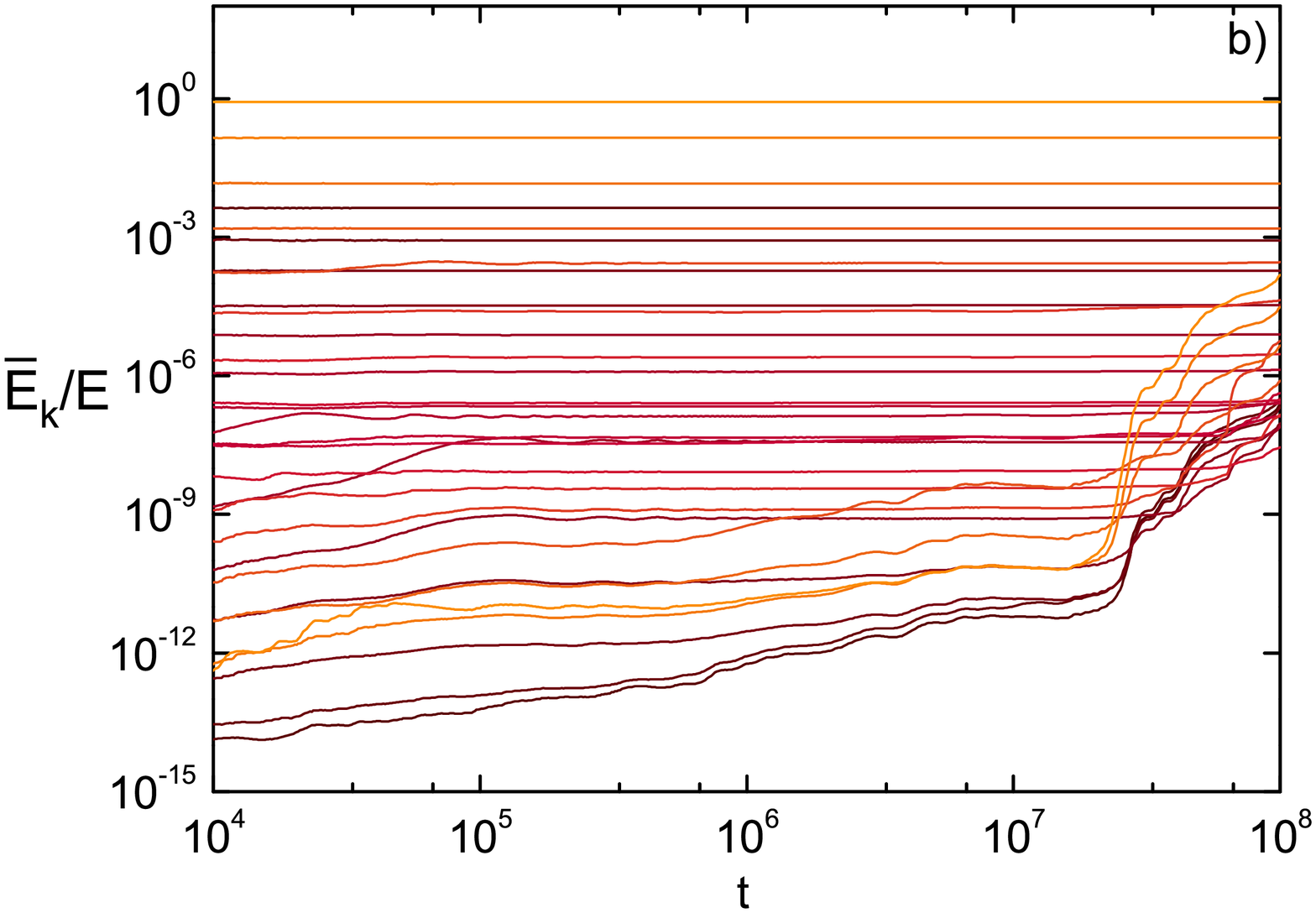}\\% Here is how to import EPS art
\includegraphics[scale=0.3 ]{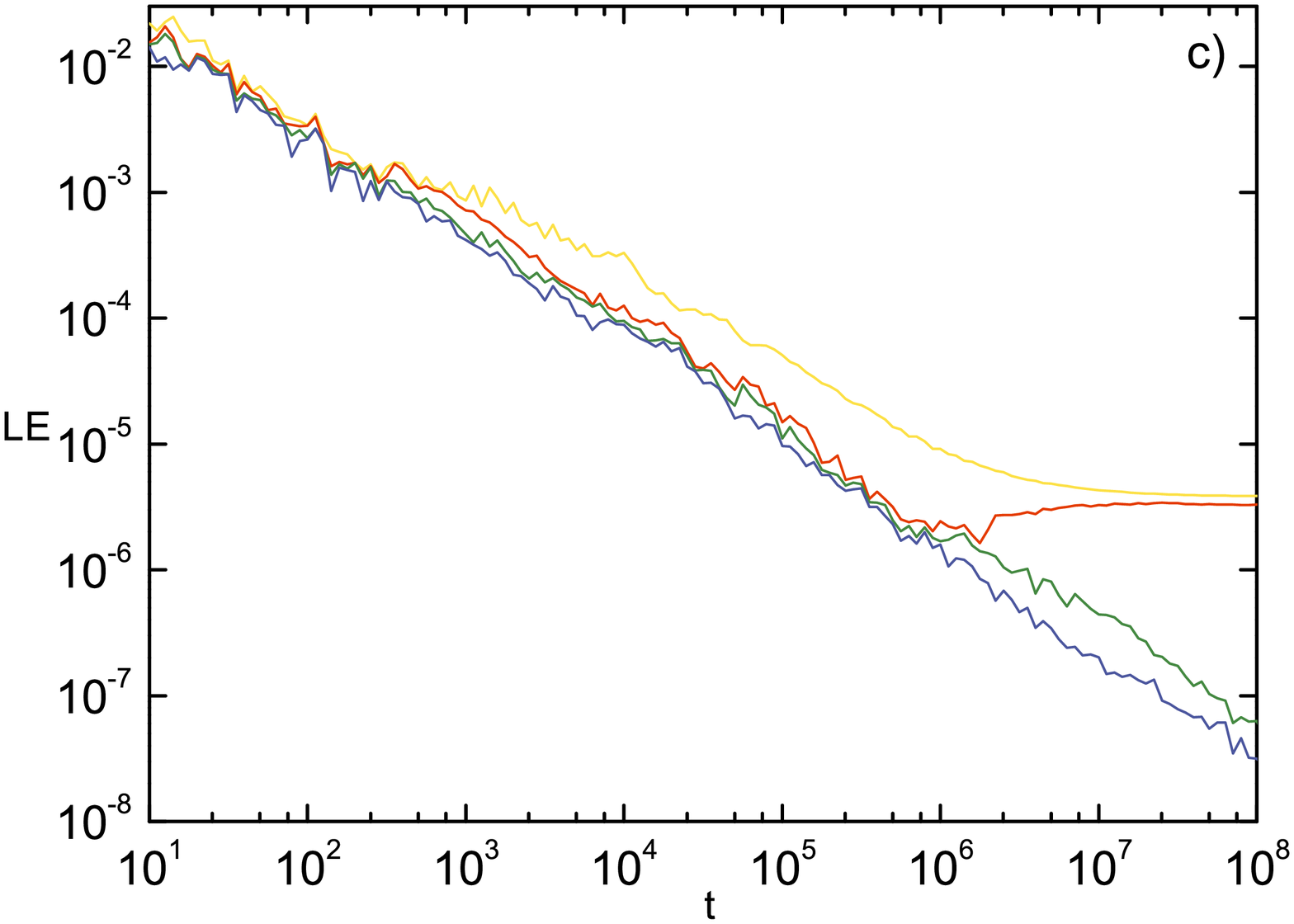}
\includegraphics[scale=0.3 ]{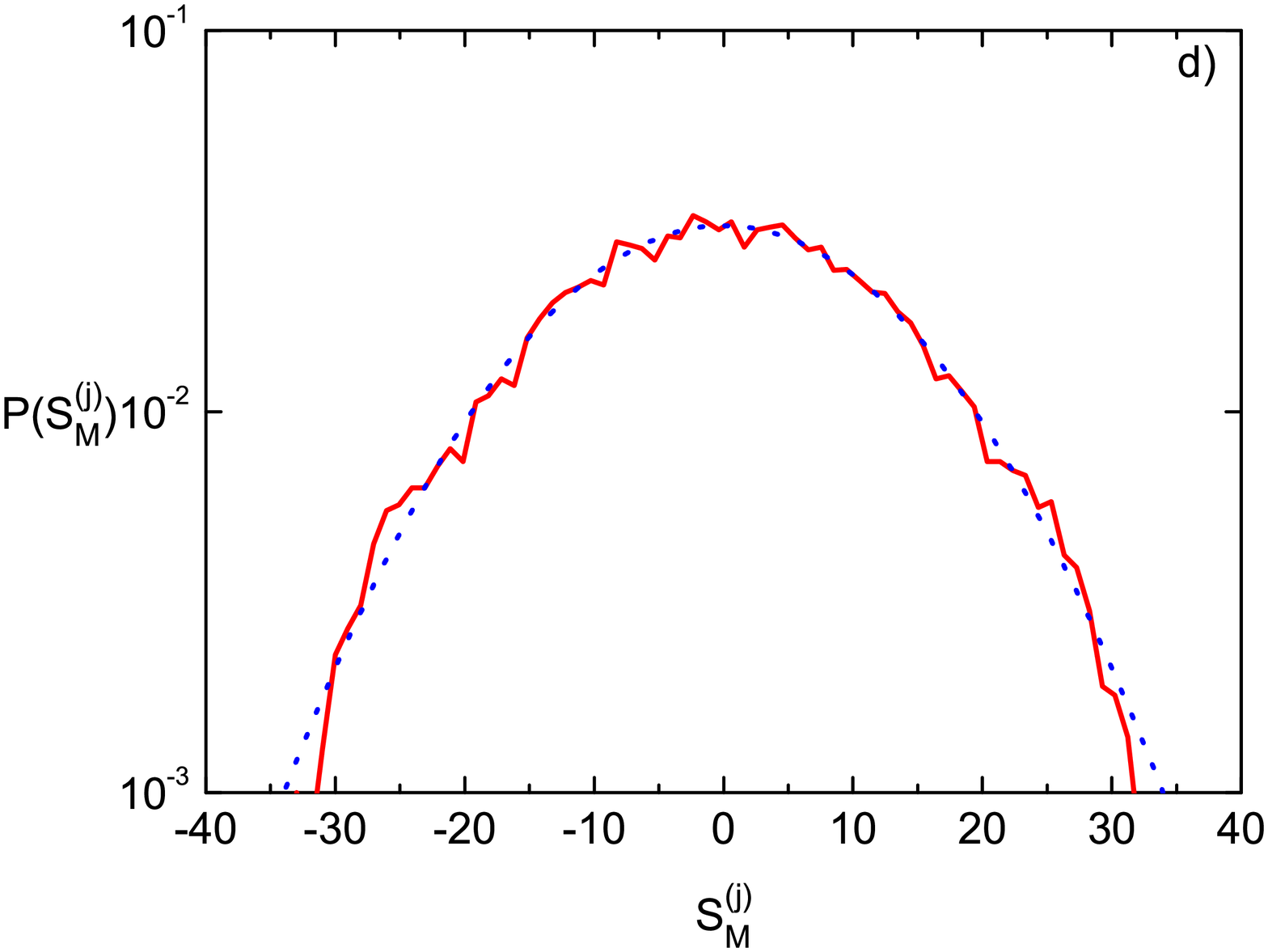}
\caption{FPU-$\alpha$ system with $N=31$, $\alpha =0.33$ and $E=4$. The initial condition here is such that all the energy is attributed to the last normal mode $k=31$. Panel a) plots the normalized averaged harmonic energy spectrum at $t_f=10^8$ showing how the total energy has been distributed among the modes. Panel b) shows the time evolution of the averaged energy spectra $\bar{E}_k/E$ of eq. \eqref{harmonicaveene}, while panel c) presents the time evolution of the four largest Lyapunov exponents. In panel d), plotting the numerical distribution (red curve) that corresponds to the final integration time $t_f=10^8$, for $N_{ic}=10^4$ time windows and $M=100$ terms in the sums, we find for the observable $\eta=q_{14}+q_{15}+q_{16}$ that the statistics is closely approximated by the Gaussian distribution ($q=1$) denoted by the blue curve and hence the motion has entered a regime of strong chaos.} \label{fig:q=31_E=4}
\end{figure}

By contrast, when we excite the $k=31$ normal mode, \textit{no diffusion} through a metastable state is observed at energies $E>0.19$, until about $E=4$ as we see in Fig. \ref{fig:q=31_E=4}. In panels a), b) and c) of that figure, we show evidence of energy transfer from the dominant modes, i.e. $N,2,N-2,4,N-4,6,N-6,\ldots$ to the tail modes with the lowest energy: $16,N-16,\ldots,30,N-30$. This transition occurs rather rapidly compared with the $k=1$ excitations, since at times up to $t_f=10^8$, energy equipartition between all modes has already taken place, as suggested by panel d) of Fig. \ref{fig:q=31_E=4}, where we have plotted in lin - log scale the numerical (red curve) and Gaussian (blue curve) distributions for this integration time and observable $\eta=q_{14}+q_{15}+q_{16}$. These results clearly show that the sum distributions of the data are very close to Gaussians ($q=1$), implying that by that time our orbit has entered a regime of strong chaos.

%%%%%%%%%%%%%%%%%%%%%%%%%%%%%%%%%%%%%%%%%%%%%%%%%%%%%%%%%%%%%%%%%%%%%%%%%%%%%%%%%%%%%%%%%%%%%%%%%%%%%%%%%%%%%%%%%%%%%%%%%%%%%%%%%%
%%%%%%%%%%%%%%%%%%%%%%%%%%%%%%%%%%%%%%%%%%%%%%%%%%%%%%%%%%%%%%%%%%%%%%%%%%%%%%%%%%%%%%%%%%%%%%%%%%%%%%%%%%%%%%%%%%%%%%%%%%%%%%%%%%

\section{Conclusions}

In this paper, we have numerically constructed pdfs of rescaled sums of observables derived from periodic orbits of the FPU-$\alpha$ Hamiltonian system. Assuming that these observables behave as independent random variables, we have attempted to determine their statistics in phase space regions of ``weak chaos'' in the context of the Central Limit Theorem. In particular, we focused our study on initial excitations of the first and last normal modes of the system and have examined the resulting dynamics from the point of view of energy diffusion processes leading to the formation of QSS, which eventually reach equipartition at sufficiently high energies and/or integration times.

At energies where the periodic orbits have turned unstable, the motion is generally expected to evolve within weakly chaotic domains, where the associated pdfs are well approximated by $q$-Gaussian distributions ($1<q<3$) for long times. These pdfs frequently represent QSS and tend to Gaussians ($q=1$) for longer integration times and/or higher energies, as the orbits diffuse into domains of ``strong chaos'' characterized by large (positive) Lyapunov exponents. 

In the case of the $k=1$ mode excitation, the energy flow to higher modes, as predicted by the theory of FPU recurrences and the breakdown of exponential localization of the energies $E_k$ was found to be connected with the appearance of a QSS whose statistics is well-described by a $q$-Gaussian with $q\approx 1.34$, at a total energy of $E=0.19$. At higher energies, e.g. $E=0.3$, this QSS is more short-lived and energy equipartition occurs more rapidly, described by $q$-Gaussian pdfs which quickly converge to Gaussians with $q=1$.  

The excitation of the last normal mode, however, yields distinctly different dynamics: At low energy values, exponential localization of linear mode energies is very dominant so that no energy diffusion to higher modes is observed and no metastable states are formed, as the motion continues to oscillate quasi-periodically on low-dimensional tori. This persists until the total energy becomes $E\approx4$, where sum distributions rapidly converge to Gaussians and energy equipartition occurs abruptly as the motion enters suddenly a regime of strong chaos.

Other authors have also studied similar QSS in conservative systems such as coupled standard maps and the so-called HMF model from the point of view of Non-Extensive Statistical Mechanics \cite{Baldovin2004a,Baldovin2004b}. In these works, the growth of variance of a temperature-like quantity, expressed by the averaged ``angular momentum'' of the models, was calculated and interesting results were obtained, but not from the viewpoint of sum distributions. The connection of such metastable states to dynamical phenomena such as the appearance of chaotic breathers and the onset of equipartition has been analyzed in detail by \cite{Cretegnyetal1998} in the vicinity of the $\pi$-mode in FPU-$\beta$ systems with periodic boundary conditions.

More recently, long-lived QSS approximated by $q$-Gaussian pdfs have been investigated in various multi-dimensional FPU-$\beta$ Hamiltonian systems in \cite{Antonopoulosetal2010}, under periodic and fixed boundary conditions. It was found that near unstable periodic orbits representing continuations of harmonic modes of the linear ($\beta=0$) problem there exist weakly chaotic domains where pdfs of the $q$-Gaussian type describe QSS for surprising long time intervals. Beyond these intervals, $q\rightarrow1$ and the pdfs quickly converge to Gaussians, as the orbits enter regimes of strong chaos and energy equipartition occurs. Finally, certain very recent results on area-preserving maps \cite{RuizBountisTsallis2010} are also in good agreement with what we have found here for multi-dimensional Hamiltonian systems. These findings suggest that the presence of chains of islands of stability in phase space and the associated diffusion and stickiness phenomena around these islands are responsible for QSS approximated by $q$-Gaussian pdfs for long times.

Nevertheless, as these authors emphasize, it is important to recall that $q$-Gaussians are \textit{not the only} possible choice for describing QSS of the type we have discussed. Indeed, in certain cases, it has been found that metastable states start as $q$-Gaussians and pass through intermediate stages where they are better approximated by other types of distributions before finally tending to Gaussians, after long enough times \cite{Antonopoulosetal2010,RuizBountisTsallis2010}. 

We believe, therefore, that the realm of conservative systems, represented by Hamiltonian flows or symplectic maps, is well suited for discovering complex metastable phenomena occurring in ``weakly chaotic'' domains and relating to more global properties of the dynamics. Even if the system is multi-dimensional, the dynamics near the boundaries of resonance islands frequently yield long-lived states with Non-Extensive statistics. Thus, we suggest that it may be possible to apply these ideas to some very slow diffusive phenomena recently observed in 1-dimensional disordered chains \cite{Flach2009,Johansson2009,Skokos2009}, where it is not yet known if, in the limit of $t\rightarrow\infty$, diffusion will extend to particles further and further away or will be hindered by the boundary of some high-dimensional KAM torus, which is believed to exist \cite{Aubry2010}.

%%%%%%%%%%%%%%%%%%%%%%%%%%%%%%%%%%%%%%%%%%%%%%%%%%%%%%%%%%%%%%%%%%%%%%%%%%%%%%%%%%%%%%%%%%%%%%%%%%%%%%%%%%%%%%%%%%%%%%%%%%%%%%%%%%
%%%%%%%%%%%%%%%%%%%%%%%%%%%%%%%%%%%%%%%%%%%%%%%%%%%%%%%%%%%%%%%%%%%%%%%%%%%%%%%%%%%%%%%%%%%%%%%%%%%%%%%%%%%%%%%%%%%%%%%%%%%%%%%%%%

\section{Appendix}
In \cite{Christodoulidi2011}, it is shown using the Poincar\'{e} - Lindstedt perturbation method, that one can construct various quasi-periodic orbits for the FPU system, under the assumption that at zeroth order we excite $s$ arbitrarily selected modes, with $1\leq s\leq N$. Actually, it is possible to prove a proposition predicting which modes will be subsequently excited in the next orders of the perturbation theory. According to this proposition, if one excites in the zeroth order of the FPU-$\alpha$ model \eqref{fpuham} the mode $k_0$, so that $s=1$ (corresponding to the q-breather solution of \cite{Flach1}), the subsequent $k_i$ modes, where $i$ denotes the order of the Poincar\'{e} - Lindstedt series, will be  
\begin{equation}
k_i= \Bigg|2\Biggr[\frac{(i+1)k_0+N}{2(N+1)}\Biggl] (N+1) - (i+1)k_0 \Bigg|,
\end{equation}
where $[\cdot]$ denotes the integer part of the argument and the $|\cdot|$ the absolute value of the argument. Thus, for $N=31$ particles, one finds the sequence reported in Sec. \ref{subsection_last_mode} for the excitable $k_0=31$-st mode.
 
%%%%%%%%%%%%%%%%%%%%%%%%%%%%%%%%%%%%%%%%%%%%%%%%%%%%%%%%%%%%%%%%%%%%%%%%%%%%%%%%%%%%%%%%%%%%%%%%%%%%%%%%%%%%%%%%%%%%%%%%%%%%%%%%%%
%%%%%%%%%%%%%%%%%%%%%%%%%%%%%%%%%%%%%%%%%%%%%%%%%%%%%%%%%%%%%%%%%%%%%%%%%%%%%%%%%%%%%%%%%%%%%%%%%%%%%%%%%%%%%%%%%%%%%%%%%%%%%%%%%%

%%%%%%%%%%%%%%%%%%%%%%%%%%%%%%%%%%%%%%%%%%%%%%%%%%%%%%%%%%%%%%%%%%%%%%%%%%%%%%%%%%%%%%%%%%%%%%%%%%%%%%%%%%%%%%%%%%%%%%%%%%%%%%%%%%
%%%%%%%%%%%%%%%%%%%%%%%%%%%%%%%%%%%%%%%%%%%%%%%%%%%%%%%%%%%%%%%%%%%%%%%%%%%%%%%%%%%%%%%%%%%%%%%%%%%%%%%%%%%%%%%%%%%%%%%%%%%%%%%%%%

\section{Acknowledgments} 

We are grateful to the organizers of the NDC2010 conference that took place at Thessaloniki during summer of 2010, in honor of Prof. T. Bountis' 60th birthday, for the opportunity they gave us to discuss our ideas with many scientists, such as Prof. Constantino Tsallis and Haris Skokos, ultimately leading to some of the results presented in this paper. We wish to thank Prof. T. Bountis for many valuable suggestions and comments on the original form and content of the manuscript. The numerical simulations presented here were performed at the high performance multi-processor computer system ``TURING'' of the University of Patras and at the computer facilities offered by the HPCS-Lab of the Technological Educational Institute of Messolonghi.

%%%%%%%%%%%%%%%%%%%%%%%%%%%%%%%%%%%%%%%%%%%%%%%%%%%%%%%%%%%%%%%%%%%%%%%%%%%%%%%%%%%%%%%%%%%%%%%%%%%%%%%%%%%%%%%%%%%%%%%%%%%%%%%%%%
%%%%%%%%%%%%%%%%%%%%%%%%%%%%%%%%%%%%%%%%%%%%%%%%%%%%%%%%%%%%%%%%%%%%%%%%%%%%%%%%%%%%%%%%%%%%%%%%%%%%%%%%%%%%%%%%%%%%%%%%%%%%%%%%%%

%%%%%%%%%%%%%%%%%%%%%%%%%%%%%%%%%%%%%%%%%%%%%%%%%%%%%%%%%%%%%%%%%%%%%%%%%%%%%%%%%%%%%%%%%%%%%%%%%%%%%%%%%%%%%%%%%%%%%%%%%%%%%%%%%%
%%%%%%%%%%%%%%%%%%%%%%%%%%%%%%%%%%%%%%%%%%%%%%%%%%%%%%%%%%%%%%%%%%%%%%%%%%%%%%%%%%%%%%%%%%%%%%%%%%%%%%%%%%%%%%%%%%%%%%%%%%%%%%%%%%

\bibliographystyle{ws-ijbc}
\bibliography{bibliography.bib}

%%%%%%%%%%%%%%%%%%%%%%%%%%%%%%%%%%%%%%%%%%%%%%%%%%%%%%%%%%%%%%%%%%%%%%%%%%%%%%%%%%%%%%%%%%%%%%%%%%%%%%%%%%%%%%%%%%%%%%%%%%%%%%%%%%
%%%%%%%%%%%%%%%%%%%%%%%%%%%%%%%%%%%%%%%%%%%%%%%%%%%%%%%%%%%%%%%%%%%%%%%%%%%%%%%%%%%%%%%%%%%%%%%%%%%%%%%%%%%%%%%%%%%%%%%%%%%%%%%%%%

\end{document}